\documentclass[fleqn,10pt]{styles/wlscirep}
\usepackage[utf8]{inputenc}
\usepackage[T1]{fontenc}
\usepackage{lineno}
\usepackage{subcaption}

\title{Generative models of MRI-derived neuroimaging features and associated dataset of 18,000 samples}
\author[1,$\dag$]{Sai Spandana Chintapalli}
\author[1,$\dag$]{Rongguang Wang}
\author[1,$\dag$]{Zhijian Yang}
\author[1,$\dag$]{Vasiliki Tassopoulou}
\author[1]{Fanyang Yu}
\author[1]{Vishnu Bashyam}
\author[1,2]{Guray Erus}
\author[3]{Pratik Chaudhari}
\author[1,4]{Haochang Shou}
\author[1,2,*]{Christos Davatzikos}
\affil[1]{Center for AI and Data Science for Integrated Diagnostics (AI2D), Perelman School of Medicine, University of Pennsylvania, Philadelphia, PA 19104, USA}
\affil[2]{Department of Radiology, Perelman School of Medicine, University of Pennsylvania, Philadelphia, PA 19104, USA}
\affil[3]{Department of Electrical and Systems Engineering, University of Pennsylvania, Philadelphia, PA 19104, USA}
\affil[4]{Department of Biostatistics, Epidemiology and Informatics, Perelman School of Medicine, University of Pennsylvania, Philadelphia, PA 19104, USA}
\affil[*]{Corresponding author: Christos Davatzikos (\href{mailto:christos.davatzikos@pennmedicine.upenn.edu}{christos.davatzikos@pennmedicine.upenn.edu})}
\affil[$\dag$]{These authors contributed equally to this work}

\begin{abstract}
Availability of large and diverse medical datasets is often challenged by privacy and data sharing restrictions. For successful application of machine learning techniques for disease diagnosis, prognosis, and precision medicine, large amounts of data are necessary for model building and optimization. To help overcome such limitations in the context of brain MRI, we present GenMIND: a collection of generative models of normative regional volumetric features derived from structural brain imaging. GenMIND models are trained on real brain imaging regional volumetric measures from the iSTAGING consortium, which encompasses over 40,000 MRI scans across 13 studies, incorporating covariates such as age, sex, and race. Leveraging GenMIND, we produce and offer 18,000 synthetic samples spanning the adult lifespan (ages 22-90 years), alongside the model's capability to generate unlimited data. Experimental results indicate that samples generated from GenMIND agree with the distributions obtained from real data. %
Most importantly, the generated normative data significantly enhance the accuracy of downstream machine learning models on tasks such as disease classification. Data and models are available at: \href{https://huggingface.co/spaces/rongguangw/GenMIND}{https://huggingface.co/spaces/rongguangw/GenMIND}.

\end{abstract}
\begin{document}
\flushbottom
\maketitle
\thispagestyle{empty}
\section*{Background \& Summary}

Empowered by widely available open datasets and challenges, in the past decade, machine learning algorithms have surpassed human-level performance on many non-trivial computer vision and natural language understanding tasks including face recognition, object detection, question answering, and translation~\cite{lecun2015deep, he2017mask, vaswani2017attention, brown2020language}.
Translating the success of artificial intelligence tools from natural image or language applications to medical domain holds immense potential for advancing automated diagnosis and precision medicine.
However, medical data usually contains sensitive patient information, which is subject to stringent privacy regulations from hospitals~\cite{chen2021synthetic, rajotte2022synthetic}. Consequently, machine learning research in medicine is contained within individual institutes or hospitals, constrained by relatively small sample sizes due to resource limitations and resulting in datasets that lack diversity in terms of patient demographics and pathology representation. 
Models trained on such less-representative medical data are subject to sample selection bias and may lack generalizability across clinical cohorts~\cite{benkarim2022population, wang2022embracing, kopal2023end, wang2023bias}. This underscores the need for collaborative efforts to access larger and more diverse medical datasets, ensuring the development of robust and trustworthy AI models in healthcare.

To address such shortages of data specifically in the context of neuroimaging, numerous international consortia have been established. Examples include ENIGMA~\cite{thompson2014enigma} and iSTAGING~\cite{habes2021brain}, which primarily compile structural, functional imaging, and genetic data from tens of thousands of individuals.
However, the usage of the data is still restricted to each consortium's participants. To broaden accessibility and enhance research potential, generative models in machine learning have garnered attention. Techniques such as kernel-density estimation (KDE)~\cite{chen2017tutorial} and generative adversarial networks (GANs)~\cite{goodfellow2014generative, yi2019generative, wang2023applications} have shown promises in synthesizing data that faithfully represents the underlying probability distributions of real-world data. This capability opens avenues for generating synthetic datasets that closely resemble actual patient data, thereby augmenting the pool available for analysis and model training. 
Recently, these models have been utilized to produce synthetic medical data and have proven valuable in many medical applications including chest X-ray screening and skin lesion detection~\cite{yu2020medical, dumont2021overcoming, dalmaz2022resvit, giuffre2023harnessing}. Consequently, these synthetic datasets can be openly shared and distributed without concerns regarding data privacy. Furthermore, having access to synthetic data that accurately represents a large population can significantly benefit downstream classification models, especially when working with a limited number of labeled examples. By leveraging synthetic data, we can bridge the gap between the available labeled samples and the diverse real-world scenarios, improving the robustness and generalization of our models.
In many studies involving  MRI (Magnetic Resonance Imaging), brain structure is commonly summarized by region-of-interest (ROI) volumes~\cite{doshi2016muse}, which are derived from structural T1-weighted MRI scans. ROI volumes are robust brain features that have been validated in many applications including disease diagnosis and prognosis~\cite{davatzikos2011prediction, rathore2017review, samper2018reproducible, wen2020convolutional, wang2023adapting}, progression modelling~\cite{marinescu2021alzheimer, tassopoulou2022deep, wijeratne2023temporal, aksman2023data, young2024data}, and pathology subtype discovery~\cite{young2018uncovering, yang2021deep, vogel2021four, yang2024gene}.

In this paper, we present GenMIND, a collection of generative models that generate normative ROI data over the adult lifespan (age range: 22 to 90 years) for different demographic groups categorized by race and sex. Furthermore, using these models, we offer a dataset of 18,000 synthetic neuroimaging samples representing a diverse global healthy adult population. Notably, we also provide the generative models alongside the dataset to enable users to customize their data synthesis. The dataset includes participants' demographic information, such as sex, age and race, which can be beneficial for research focusing on mitigating algorithmic bias and promoting fairness.
Our approach to generating synthetic data involved training generative models, specifically KDE models, on 34,000 subjects from the iSTAGING consortium~\cite{habes2021brain} to synthesize both brain anatomical ROI volumes and their associated demographics.
In the experiments, we assess the quality of the synthetic data from both qualitative and quantitative perspectives.
For example, we create plots highlighting the similarity between the real and synthetic data distributions across all ages groups.
In terms of quantitative evaluation, we use statistical tests and machine learning models to show the similarity in distributions between synthetic and real data. We also examine the fidelity of the synthetic data by performing covariates prediction including sex or race classification and age regression tasks. Finally, we show the utility of the synthetic data in real world applications such as brain age gap prediction~\cite{franke2019ten}.

GenMIND, a comprehensive structural brain imaging generator and dataset, aims to significantly contribute to advancing machine learning research within the field of neuroimaging. In particular, GenMIND can facilitate:
1) Local Disease Population Comparisons: Researchers can leverage GenMIND to compare their own patient populations to GenMIND's normative dataset. To ensure robust inference that is invariant to inter-site differences, researchers should first use part of their control data to harmonize their measures with those of GenMIND.
2) Brain Age Prediction Models: GenMIND serves as a resource for training brain age prediction models. These models have been used to detect the effects of neurodegenerative and neuropsychiatric disorders on the brain. In the technical validation section, experiments using GenMIND have explored the relationship between brain age residuals and cognitive scores.Beyond brain age prediction, the synthetic data generated by GenMIND can also be leveraged to train other machine learning models and adapt them for smaller-scale studies through techniques like transfer learning or domain adaptation.
3) Enriching Healthy Controls in Classification Studies: The dataset enhances the healthy control class for discriminative analysis and disease subtyping research. For instance, in an Alzheimer’s disease diagnosis experiment, GenMIND has been utilized for data augmentation, leading to more robust results.
4) Synthetic Data Generator Models: Beyond providing the GenMIND dataset containing 18,000 samples, researchers also have access to the synthetic data generator models. These pre-trained models allow customization of brain volume synthesis based on factors such as sex, age range, and race groups.  

Our goal is to continue expanding GenMIND with additional covariates, including genetic risk factors, cognitive scores, and biomarker data. This expansion will enable users to enrich their own datasets by synthesizing highly specific brain ROI measures tailored to their individual studies.

\begin{table}[!t]
\caption{Synthetic data table comprising of brain ROI volumes along with demographic information. Sex variable is coded as F for female and M for male.}
\label{tab:1}
\centering
\begin{tabular}{|l|l|l|l|l|l|l|l|}
\hline
SampleID &Sex & Race & Age &  3rd Ventricle & 4th Ventricle & Right Accumbens Area & ...
\\
\hline
Synth1 & F&	Black&	83.49359824	&572.7620447&	1298.91257&	438.5023773&...\\
\hline
Synth2& M & White	&62.13155172&	1204.073174	&1149.880788	&246.0560264&...\\
\hline
.&.&.&.&.&.&.&.\\
.&.&.&.&.&.&.&.\\
.&.&.&.&.&.&.&.\\
\hline
\end{tabular}
\end{table}


\section*{Methods}
In the following subsections we describe i) the real dataset used to construct the GenMIND generative model, and ii)  the approach used for training the generative model for synthetic data generation.

\paragraph{Real data}
For our real data, we use the iSTAGING consortium~\cite{habes2021brain} that consolidated and harmonized imaging and clinical data from multiple cohorts spanning a wide age range (22 to 90 years). Our data consists of multimodal neuroimaging and demographic measures taken from  subjects labeled as cognitively normal in the iSTAGING consortium. Specifically, the neuroimaging measures are the 145 anatomical brain ROI volumes (119 ROIs in gray matter, 20 ROIs in white matter and 6 ROIs in ventricles) from baseline scans extracted using a multi‐atlas label fusion method~\cite{doshi2016muse}. To mitigate site effects, ComBat-GAM harmonization~\cite{pomponio2020harmonization} was applied to these 145 ROI volumes while accounting for age, sex and intracranial volume (ICV). The demographic measures including subjects' age, sex, and race were accounted in the synthesis.  
Employing a stratified approach, subjects in the real data were grouped based on race and sex covariates, resulting in six subsets or categories: white male, white female, black male, black female, asian male, and asian female. Before fitting the generative model, within each category, the feature vector of 145 ROI volumes along with the age variable is normalized using the mean and the standard deviation. Mean and standard deviation values for each category were retained to facilitate the back transformation of synthesized data from normalized space to the original space.

\paragraph{Generative model training}
We employ a separate non-parametric kernel density estimation (KDE) model for each category, using a Gaussian kernel to delineate the joint probability density of age and the 145 ROI volumes. The selection of the Gaussian kernel is crucial; its smooth, bell-shaped profile is well-suited for generating a continuous representation of the underlying distribution from discrete observations. Within this framework, each data point is represented by a local Gaussian density surface centered at the observed location. Collectively, the overall density is smoothly estimated using composite functions of these local densities, based on the discrete empirical data. This approach facilitates the estimation of the multivariate joint probability density function for age and ROI volumes within each distinct category.
Generating new synthetic data involves a two-step process: model fitting and sampling. Initially, a multivariate KDE model is fitted to the real data of each sex and race combination with carefully tuned bandwidth parameters. We then generate new data points by sampling from this refined distribution. Each synthesized vector is of size 146; the first 145 elements correspond to the ROI volumes, and the last element represents age. The sampling process introduces randomness, ensuring that while the synthetic data are not exact replicas of the original, they adhere to the same statistical properties.

\paragraph{Hyper-parameter selection}
 The bandwidth parameter determines the width of the Gaussian kernel and plays a crucial role in KDE. A smaller bandwidth results in a more granular estimate with greater detail, reflecting minor variations in the data distribution, whereas a larger bandwidth leads to a smoother and more generalizable estimate. By carefully selecting the bandwidth, the KDE model can effectively navigate the trade-off between overfitting, where the model captures excessive noise, and underfitting, where important features of the data distribution are overlooked. To identify the optimal bandwidth for the kernel in Kernel Density Estimation (KDE), we conducted a grid search spanning bandwidths from 0.5 to 1. The selection criterion for the optimal bandwidth was the maximization of the log-likelihood score across these values. For the majority of the categories analyzed, the optimal bandwidth converged at approximately 0.7. The search space of the bandwidth is constrained to 0.5 to 1 because our dataset is normalized ($\sim N(0,1)$). A small bandwidth will overfit the training data, since it is going to fit a narrow Gaussian kernel around any point; whereas a large bandwidth, close to 1, will result in an overly smoothed density estimate that fails to maintain the correlation between age and the 145 ROIs. The model implementation and the bandwidth grid-search are carried out using the scikit-learn library~\cite{scikit-learn}. 

\section*{Data Records}

The GenMIND dataset and model are hosted at \href{https://huggingface.co/spaces/rongguangw/GenMIND}{https://huggingface.co/spaces/rongguangw/GenMIND}.
The "genmind\_dataset.csv" file contains 18,000 synthetic samples with 3000 samples allocated per combination of race and sex, a deliberate sampling strategy aimed at ensuring thorough representation across demographic groups. Each sample comprises of 148 features (145 brain ROI volumes along with age, sex, and race information). See Table.~\ref{tab:1} for an example of the synthetic dataset. We enhance the utility of our dataset by providing users with access to both the synthetic data and the fitted Kernel Density Estimation (KDE) models. This provision equips researchers with the ability to generate synthetic data that aligns precisely with their research requirements.

\section*{Technical Validation}
\begin{figure}[!t]
\centering
\begin{subfigure}{1\textwidth}
  \centering
  \includegraphics[width=0.8\linewidth]{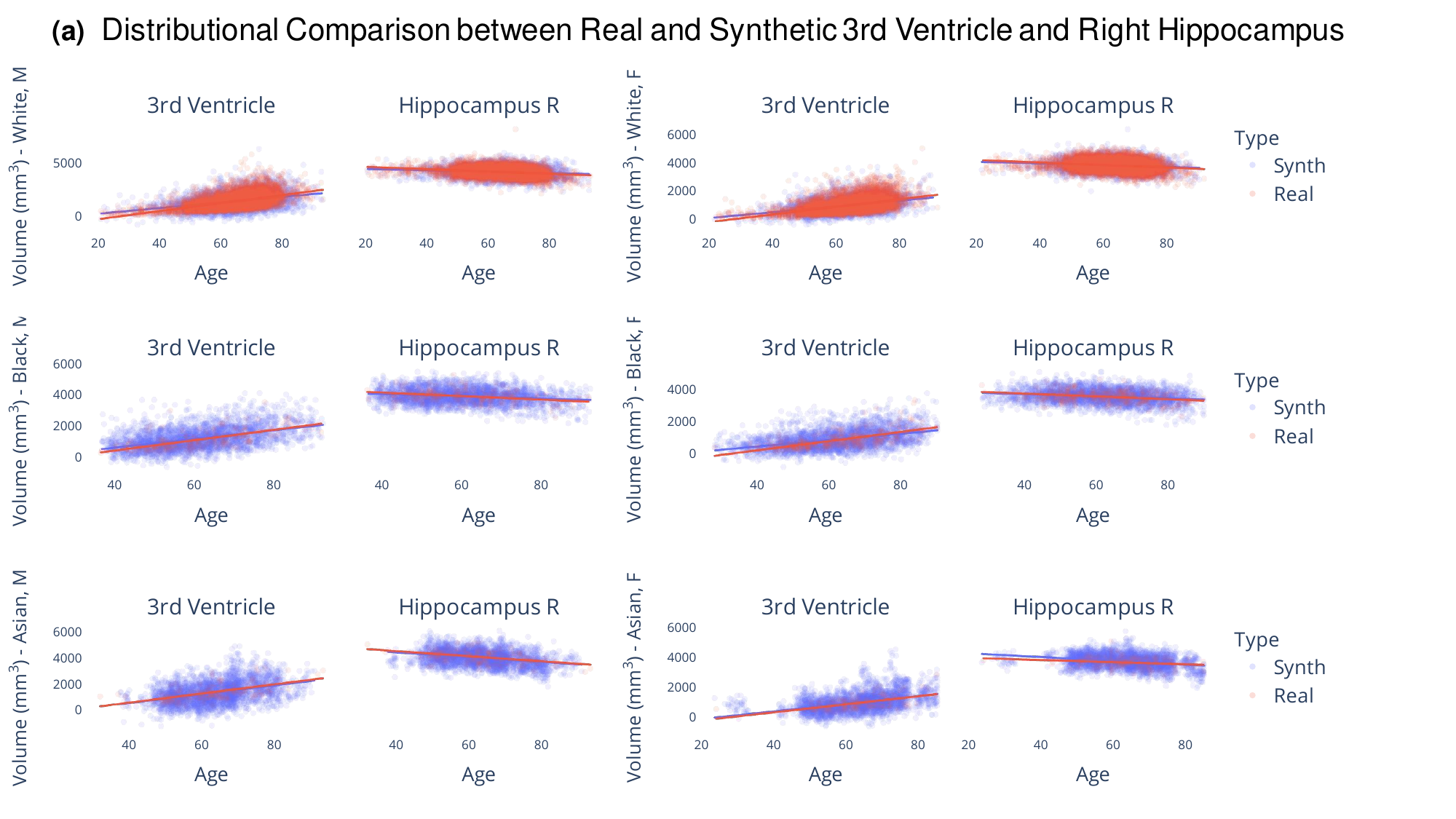}
  \label{fig1a}
\end{subfigure}
\hfill
\begin{subfigure}{0.8\textwidth}
\centering
  \includegraphics[width=1\linewidth]{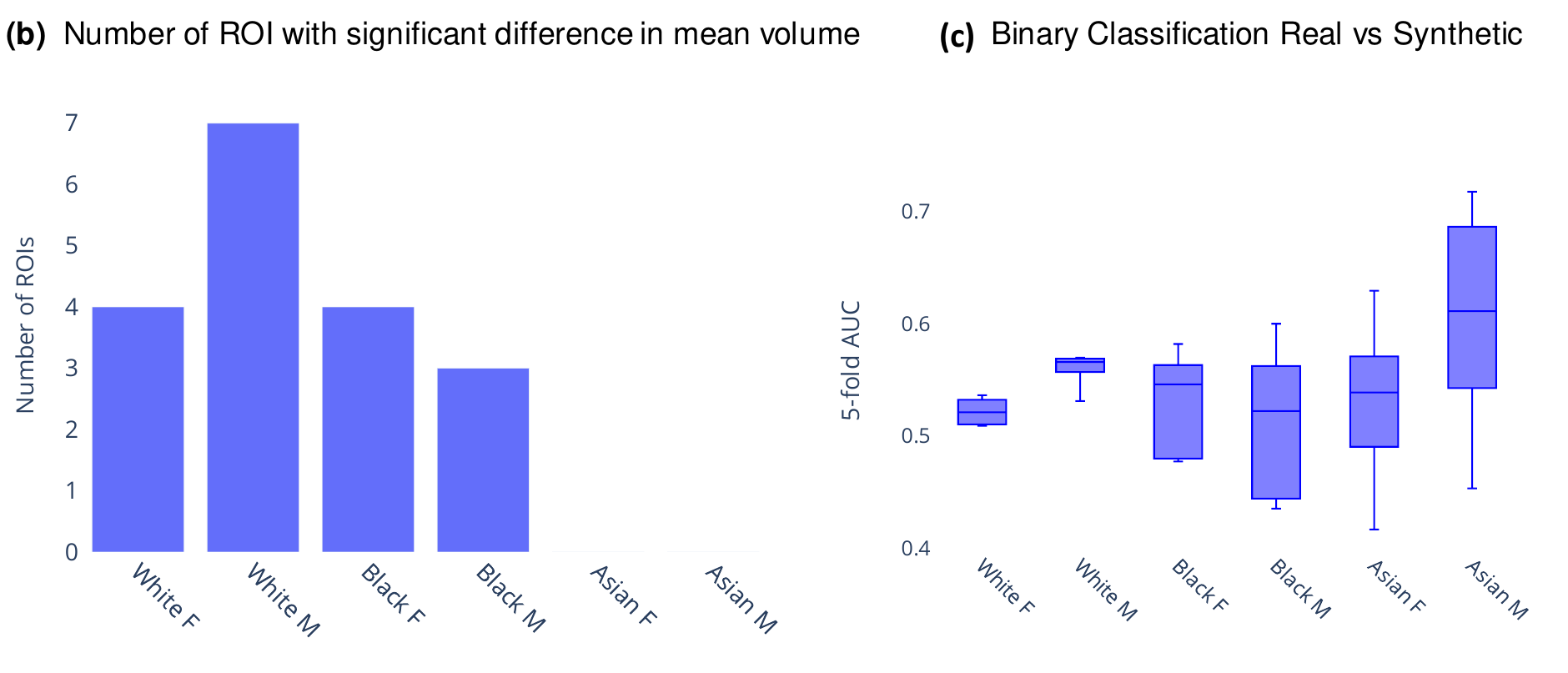}
  \label{fig1b}
\end{subfigure}
\caption{Comparing synthetic and real data distributions. (a) Qualitative assessment of real versus synthetic ROI volume data distributions in the 3rd ventricle and right hippocampus. (b) Results from univariate statistical analysis to check for differences in synthetic and real ROI volumes. (c) 5-fold cross-validated AUC scores for binary classification of real and synthetic data using support vector machine regression indicate that the generated dataset is indistinguishable from the real dataset.}
\label{fig1}
\end{figure}


We conduct a comprehensive validation of the synthetic data through two levels of analysis. In the first level, we employ a combination of statistical and machine learning techniques to ensure that the synthetically generated brain ROI volume data closely mirrors the distribution observed in real ROI volume data across demographic variables such as age, sex, and race.  Subsequently, at the second level, we aim to validate the practical use of synthetic data in real-world scientific applications, such as brain age gap estimation and data augmentation. Through these rigorous analyses, we aim to ascertain the accuracy, reliability, and utility of synthetic data in both replicating real data distributions and facilitating meaningful applications in scientific research and clinical practice.

\subsection*{Assessing Fidelity: Statistical and Machine Learning Analysis Comparing Synthetic and Real Data Distributions}
In this section, we examine synthetic samples generated from the KDE models in comparison with a held-out dataset containing real samples from the iSTAGING consortium \cite{habes2021brain}.
Given that the synthetic data spans the entire adult lifespan and covers all combinations of race and sex, Fig.\textcolor{blue}{1a} serves to visually illustrate the similarity in distributions between real and synthetic data across several ROIs. From Fig. \textcolor{blue}{1a} it is evident that for these two representative examples, i.e. 3rd ventricle and right hippocampus, there is significant overlap between the generated data samples and the real data samples. Furthermore, the age trend observed in the synthetic data roughly mirrors that observed in the real data. To further evaluate the fidelity of the ROI volume distributions, univariate statistical analysis was conducted via linear regression of each ROI volume on group (real versus synthetic) while adjusting for age as a covariate  ~\cite{dismuke2006ordinary}. Bonferroni correction~\cite{bland1995multiple} was applied to adjust for multiple comparisons, resulting in a significance threshold of $\alpha = \frac{0.05}{145} = 0.0003$. Fig. \textcolor{blue}{1b} presents the count of ROIs, out of the total 146 ROIs, with statistically significant differences in group means for every combination of race and sex.

Additionally, to assess the quality of the multivariate ROI volume distributions, we use the support vector machine (SVM)~\cite{hearst1998support} classifier that is trained to  differentiate between synthetic and real samples. An area under the receiver operating characteristic curve (AUC)~\cite{pepe2006combining} of 0.5 would suggest that the classifier is making random predictions and cannot distinguish between the two groups. Therefore, if the SVM fails to differentiate synthetic samples from real samples, we can infer that the generated dataset is indistinguishable from the real dataset.  Fig. \textcolor{blue}{1c} summarizes the findings across different race and sex categories. While there remains some classification capability (AUC > 0.5 in white males, white and black females), suggesting slight differences between the two distributions, subsequent analyses under practical applicability section demonstrate that this discrepancy does not confound or significantly impact the utility of the synthetic dataset.

\begin{figure}[!t]
\begin{center}
\includegraphics[height=9cm]{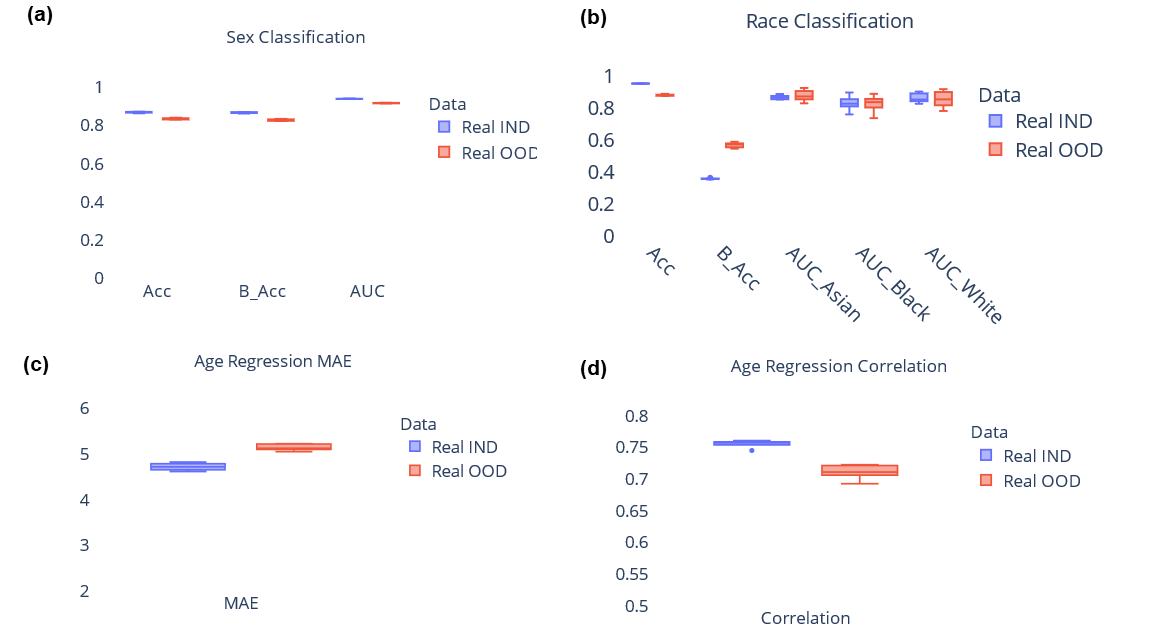}
\end{center}
\caption 
{ \label{fig:agesexrace_all}
Comparison between model trained on real data and model
trained on synthetic data for covariate prediction. Results from (a) sex classification, (b) race classification, and (c), (d) age regression. Results are shown from 5-fold cross validation. Real IND refers to results from model trained on real data and tested on held-out real data, Real OOD refers to results from model trained on synthetic data and tested on held-out real data. Acc: Accuracy, B\_Acc: Balanced Accuracy, AUC: Area under receiver operating characteristic curve, MAE: Mean absolute error.} 
\end{figure}

Since the data for each combination of race and sex was generated from different KDE models, it is critical to assess whether the relationship between ROI volumes and covariates (age, sex, race) in the synthetic data aligns with observations from real data. To carry out this analysis, we build machine learning models  for covariate prediction, specifically we use gradient boosted trees (implemented using XGBoost library~\cite{chen2016xgboost}). These models are trained to predict covariates based on ROI volume data, performing tasks such as age regression, sex classification, and race classification. By comparing the performance of the models trained on synthetic data with those trained on the real data, we can determine if synthetic data is a suitable substitute for real data. Sex and race classification performance is evaluated using metrics such as accuracy, balanced accuracy, and AUC. Age regression performance is evaluated using mean absolute error (MAE) and Pearson's correlation~\cite{sedgwick2012pearson} between predicted age and ground truth age. Fig.\ref{fig:agesexrace_all} shows the results for covariate prediction, we used 5-fold cross-validation during analysis. While models trained on real data generally outperform those trained on synthetic data, it is noteworthy that the performance of models trained on synthetic data is comparable to their real data counterparts. These results suggest that synthetic data can serve as a valuable alternative to real data, when the latter is not available. The next section delves into the practical applicability of synthetic data, exploring its potential benefits and limitations in real-world scenarios.



\subsection*{Practical Applicability}
In addition to validating the quality of the generated data, we conducted comprehensive assessments to showcase the practical applications of synthetic data across various scenarios. This included its efficacy in augmenting training datasets for disease classification and deriving clinically meaningful estimates of brain age gaps. In this section, we utilized the ADNI study as a held-out dataset to assess the aforementioned properties. The synthetic data was derived from a KDE model retrained on the remaining studies, following the steps outlined in Methods.

\subsubsection*{Data augmentation}
Deep learning methods usually require a large number of training samples, which are laborious and costly to obtain, especially for brain MRI studies. Moreover, datasets focusing on specific diseases may sometimes lack a sufficient number of healthy controls. To address this, we investigated the feasibility of using synthetic data to supplement the normal control (CN) group. We centered our analysis on the mild cognitive impairment (MCI) or Alzheimer's disease (AD) classifications. Leveraging data from the ADNI study~\cite{jack2008alzheimer}, we evaluated whether synthetic data based data augmentation can help improve classification performance. We divided the ADNI dataset, allocating 500 cognitively normal participants for training, 368 for testing, and evenly dividing 1101 MCI and 419 AD participants between training and testing sets. To ensure the robustness of our findings, we conducted fifty distinct random splits.
For both CN vs MCI and CN vs AD classifications, we trained the SVM algorithm on the training set and assessed its performance on the test set using the AUC metric. Notably, the training sets comprised different proportions of CN data from both real and synthetic datasets, allowing us to evaluate the specific contribution of synthetic data to performance enhancement (Fig. 3).\\
Figure 3 illustrates that a reduced number of training samples from the normal control group (500 v.s. 100) results in lower MCI and AD classification performances, underscoring the importance of augmenting the healthy control dataset. Supplementation of the CN set with synthetic normal control data progressively improved classification AUCs, although the rate of improvement slowed as the proportion of synthetic data within the CN set increased.

\begin{figure}[!t]
\begin{center}
\includegraphics[height=6cm]{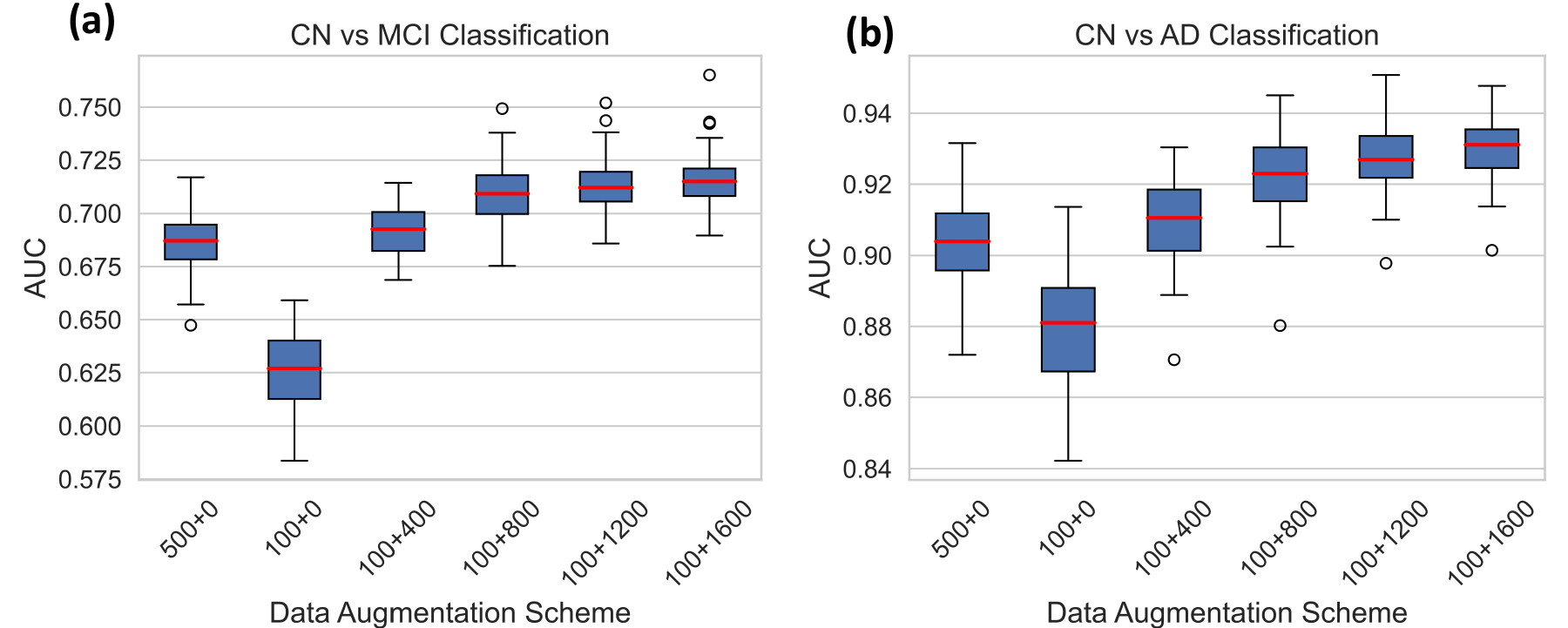}
\end{center}
\caption{AUC scores for (a) CN vs MCI and (b) CN vs AD classifications using different data augmentation schemes. The X-axis lists out the different schemes, for instance 100+400 describes experiment where model was trained using a control data set comprising of 100 real samples and 400 synthetic samples. The classification process was performed 50 times for each augmentation scheme, using fifty random splits of training and testing sets.}
{ \label{fig:data_augmentation}} 
\end{figure}

\subsubsection*{Brain age gap estimation}
Brain age gap\cite{smith2019estimation, franke2019ten, jonsson2019brain, leonardsen2022deep, zhou2023multiscale}, the difference between predicted brain age and actual chronological age, indicates deviations from normal brain aging and proves important for assessing neurological health. Utilizing large-scale synthetic control data can potentially enhance the development of age-prediction models, offering more reliable and clinically relevant brain age gap estimations. We trained XGBoost regression models on both synthetic and real control data, and used the trained models for calculating brain age gaps for CN and MCI/AD participants in the held-out ADNI dataset. We will refer to brain age gaps derived using the synthetic data model as OOD brain age gaps and refer to brain age gaps derived using the real data model as IND brain age gaps. To mitigate biases inherent in predicted brain age estimates, we applied Cole’s method \cite{zhang2023age} before calculating the brain age gaps. Further, to examine their effectiveness in indicating cognitive decline and underlying neuropathology, we examined their correlations with Mini Mental State Examination (MMSE) scores — a widely utilized cognitive assessment for measuring cognitive impairment.

As shown in Fig. 4, both OOD and IND brain age gaps show no significant correlations with MMSE among CN participants. However, among MCI/AD participants, they exhibit significant correlations ($p<0.0001$). Interestingly, the Pearson's correlation coefficient for OOD brain age gaps ($\rho=-0.235$) is similar to that observed for IND brain age gaps($\rho= -0.304 $). This further underscores the potential of regression models trained on extensive synthetic datasets to provide brain age gap estimation with increased clinical significance.

\begin{figure}[!t]
\begin{center}
\includegraphics[height=10cm]{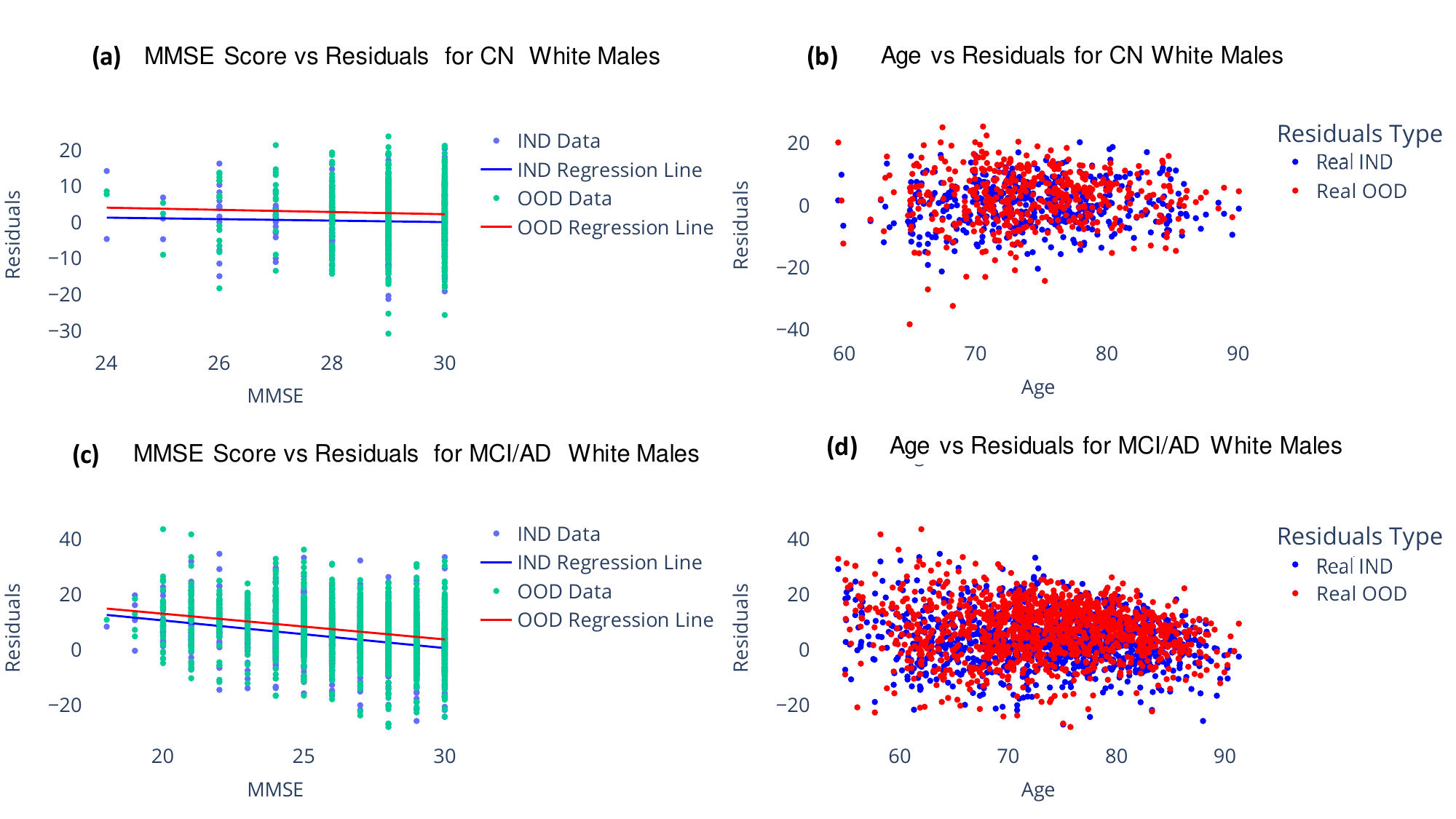}
\end{center}
\caption 
{ \label{fig:brain_age_gap}
Comparison of brain age gap for ADNI participants estimated using a model trained on real data and another model trained on synthetic data. (a) and (b) show the bias corrected brain age gap versus MMSE scores for subjects labelled as CN and MCI/AD respectively. (c) and (d) show the bias corrected brain age gap versus Age for subjects labelled as CN and MCI/AD respectively. Real IND refers to results from model trained on real data and tested on held-out real data, Real OOD refers to results from model trained on synthetic data and tested on held-out real data.} 
\end{figure} 
\section*{Usage Notes}

By employing rigorous statistical and machine learning analyses, we have demonstrated that the synthetic data generated by GenMIND closely aligns with real data distributions across various demographic variables, including age, sex, and race. Our evaluations, detailed in the technical validation section, indicate that the multivariate distributions learned by the model accurately preserve covariate effects, ensuring that the synthetic data maintains the integrity of the original data's demographic characteristics. The practical utility of GenMIND has been illustrated through various applications, showing its potential to serve as either a substitute or complement to real data. In scenarios with class imbalances in real datasets, we have shown how researchers can expand their sample sizes with synthetic data and improve the performance of machine learning models. We have also show the utility of GenMIND in brain age predictions. As it is designed to emulate a reference dataset representing a global healthy population across the human lifespan, researchers can leverage GenMIND to compare their small cohorts to the reference population, gaining insights into age-related changes and deviations from expected brain development trajectories, thus enhancing our understanding of neurodevelopmental processes and age-related neurodegeneration.\\
Moreover, synthetic data holds significant potential for harmonization efforts in integrating and comparing datasets from diverse sources. Synthetic data can be specifically tailored to generate covariate-matched control data, ensuring that external studies can be effectively harmonized with existing datasets. This approach enables researchers to create control groups with similar demographic characteristics as that of their study population, thereby reducing confounding effects and then utilizing a harmonization technique \cite{horng2022improved} to align their external data to the synthetic matched control data. This enhances the robustness of comparative analyses.\\ 
While we have demonstrated the potential uses of synthetic data, there are certain limitations that end-users should consider when using GenMIND. The quality of synthetic samples is influenced by the selection of the bandwidth parameter in KDE, and although the synthetic data statistically resembles real data, it may contain noise due to kernel smoothing. This noise might affect the results of any analysis performed with GenMIND, so users should interpret their results with caution. Additionally, when integrating their own data with GenMIND, users should be aware of potential site-related differences that might impact their analyses. There are techniques available that can help the user harmonize their data to GenMIND\cite{hu2023image} to mitigate site-related effects and facilitate more robust analyses.\\
In conclusion, GenMIND represents an advancement in the generation and application of synthetic neuroimaging data, providing a valuable resource for the neuroimaging community. By continuing to expand GenMIND with additional covariates, including genetic risk factors, cognitive scores, and biomarker data, we aim to further enhance its utility and applicability in diverse research contexts. This work underscores the importance of synthetic data in advancing neuroimaging research, promoting data accessibility, and ensuring the development of robust, generalizable machine learning models in healthcare.

\section*{Code availability}
All our model building and analysis were carried out in python. All our data transformations are described in detail in the methods section. Statistical analyses were conducted via online python packages, statsmodels 0.8.0, SciPy 1.6.3, NumPy 1.16.6 and pandas 0.21.0. Our machine learning experiments were conducted with version 1.1.3 of the Scikit-learn library\footnote{https://scikit-learn.org}. The trained KDE models are available at \href{https://huggingface.co/spaces/rongguangw/GenMIND}{https://huggingface.co/spaces/rongguangw/GenMIND}. Sample code for model training and data generation is available at \href{https://huggingface.co/spaces/rongguangw/GenMIND/blob/main/script/synthetic_data_generation.ipynb}{https://huggingface.co/spaces/rongguangw/GenMIND/blob/main/script/synthetic\_data\_generation.ipynb}.

\bibliography{refs}

\section*{Acknowledgements}
The iSTAGING consortium is a multi-institutional effort funded by the National Institute on Aging with grant number RF1AG054409. Data used in preparation of this article were in part obtained from the Alzheimer’s Disease Neuroimaging Initiative (ADNI) database (adni.loni.usc.edu). As such, the investigators within the ADNI contributed to the design and implementation of ADNI and/or provided data but did not participate in analysis or writing of this report. A complete listing of ADNI investigators can be found at: https://adni.loni.usc.edu/wpcontent/uploads/how\_to\_apply/ADNI\_Acknowledgement\_List.pdf. 
ADNI is funded by the National Institute on Aging, the National Institute of Biomedical Imaging and Bioengineering, and through generous contributions from the following: AbbVie, Alzheimer’s Association; Alzheimer’s Drug Discovery Foundation; Araclon Biotech; BioClinica, Inc.; Biogen; Bristol-Myers Squibb Company; CereSpir, Inc.; Cogstate; Eisai Inc.; Elan Pharmaceuticals, Inc.; Eli Lilly and Company; EuroImmun; F. Hoffmann-La Roche Ltd and its affiliated company Genentech, Inc.; Fujirebio; GE Healthcare; IXICO Ltd.; Janssen Alzheimer Immunotherapy Research \& Development, LLC.; Johnson \& Johnson Pharmaceutical Research \& Development LLC.; Lumosity; Lundbeck; Merck \& Co., Inc.; Meso Scale Diagnostics, LLC.; NeuroRx Research; Neurotrack Technologies; Novartis Pharmaceuticals Corporation; Pfizer Inc.; Piramal Imaging; Servier; Takeda Pharmaceutical Company; and Transition Therapeutics. The Canadian Institutes of Health Research is providing funds to support ADNI clinical sites in Canada. Private sector contributions are facilitated by the Foundation for the National Institutes of Health (www.fnih.org). The grantee organization is the Northern California Institute for Research and Education, and the study is coordinated by the Alzheimer’s Therapeutic Research Institute at the University of Southern California. ADNI data are disseminated by the Laboratory for Neuro Imaging at the University of Southern California.

\section*{Author contributions statement}
Conceptualization, Methodology, Formal Analysis, Investigation \& Writing: Sai Spandana Chintapalli, Rongguang Wang, Zhijian Yang, and Vasiliki Tassopoulou; Conceptualization, Methodology \& Formal Analysis: Fanyang Yu and Vishnu Bashyam; Conceptualization: Guray Erus; Conceptualization \& Review: Pratik Chaudhari; Review, Editing \& Supervision: Haochang Shou and Christos Davatzikos.

\section*{Competing interests}
The authors declare no competing interests.

\end{document}